\documentclass[a4paper,11pt]{article}

\usepackage{listings,multicol,multirow}
\usepackage{fullpage}
\usepackage{url}
\usepackage[utf8]{inputenc}
\usepackage[english]{babel}
\usepackage{amsmath}
\usepackage{amsfonts}
\usepackage{amssymb}
\usepackage{amsthm}
\usepackage[dvipsnames]{xcolor}
\usepackage{hyperref}
\usepackage{comment} 
\usepackage{todonotes}
\usepackage{graphicx}
\usepackage{xspace}
\usepackage{paralist}
\usepackage[margin=1cm]{caption}
\usepackage{pifont}
\usepackage{authblk}

\usepackage{paralist}

\usepackage[lambda,
 advantage,
 operators,
 sets,
 adversary,
 landau,
 probability,
 notions,
 logic,
 ff,
 mm,
 primitives,
 events,
 complexity,
 asymptotics,
 keys]{cryptocode}

\theoremstyle{plain}
\newtheorem{theorem}{Theorem}[section]
\newtheorem{lemma}[theorem]{Lemma}

\newtheorem{definition}{Definition}[section]

\DeclareMathAlphabet{\pazocal}{OMS}{zplm}{m}{n}
\DeclareMathAlphabet{\mathbfcal}{OMS}{cmsy}{b}{n}

\newcommand{\Com}{\textsf{Com}}
\newcommand{\Ver}{\textsf{Ver}}

\newcommand{\Share}{\textsf{Share}\xspace}

\newcommand{\EasyCrypt}{\textsf{EasyCrypt}\xspace}
\newcommand{\CertiCrypt}{\textsf{CertiCrypt}\xspace}
\newcommand{\CryptHOL}{\textsf{CryptHOL}\xspace}
\newcommand{\Coq}{\textsf{Coq}\xspace}

\newcommand{\OCaml}{\textsf{OCaml}\xspace}

\newcommand{\V}{\mathcal{V}}

\newcommand{\result}{\mathsf{res}}

\newcommand{\ec}[1]{\lstinline[mathescape,language=easycrypt,xleftmargin=0pt,xrightmargin=0pt,style=easycrypt-pretty,mathescape]{#1}}

\lstdefinelanguage{easycrypt}{
  style=easycrypt-default,
  keywordsprefix={'},
  morekeywords=[1]{unit,bool,int,real,list,matrix,option,distr,fset,array},
  morekeywords=[2]{type,op,axiom,lemma,module,pred,const,declare},
  morekeywords=[3]{var,proc,let,return},
  morekeywords=[4]{while,if,then,else,fun},
  morekeywords=[5]{theory,end,clone,import,export,as,section,of,proving,with},
  morekeywords=[6]{forall,exists},
  morekeywords=[7]{idtac,change,beta,iota,zeta,logic,delta,simplify,congr,generalize,
                   pose,split,left,right,case,intros,cut,elim,apply,rewrite,elimT,subst,
                   progress,trivial},
  morekeywords=[8]{by,assumption,smt,reflexivity},
  morekeywords=[9]{first,last,do,try},
  morekeywords=[10]{circuit_t,wire_t,wid_t,gid_t,topology_t,gates_t,pinput_t,sinput_t,poutput_t,rand_t,party_pdata_t,party_sdata_t,party_sdata_tuple_t,pid_t,pid_mpc_t,gate_rand_t,gate_rands_t,rands_t,messages_t,gate_local_trace_t,in_messages_t,out_messages_t,trace_t,gate_traces_t,traces_t,view_t,views_t,input_t,inputs_t,output_t,outputs_t,leakage_t,message_t,messages_tuple_t,leakages_t,secret_t,share_t,shares_t,msg_t,commitment_t,opening_string_t,commit_info_t,input1_t,input2_t,rand1_t,rand2_t,output1_t,output2_t,leakage1_t,leakage2_t,witness_t,statement_t,prover_input_t,verifier_input_t,prover_rand_t,verifier_rand_t,prover_output_t,verifier_output_t,prover_leakage_t,verifier_leakage_t,challenge_t,response_t,prover_rand_t,verifier_rand_t,prover_state_t,verifier_state_t,pid_mpc_t,pid_zk_t,committed_view_t,committed_views_t,msgs_t,randg_t},
  morekeywords=[11]{Rand_t,Distinguisher_t,Evaluator_t,Simulator_t,FRand_t,Adversary_t,MProver_t,MVerifier_t,RandP_t,RandV_t},
  morekeywords=[12]{Idle,api_input,api_output,api_operator,API,global_state,AStep},
  morekeywords=[13]{group,t,skey,pkey,plaintext,ciphertext,msg,cipher,Scheme,Adv},
  morecomment=[n][\itshape\color{darkgray}]{(*}{*)},
  morecomment=[n][\bfseries\color{darkgray}]{(**}{*)}
}

\lstdefinestyle{easycrypt-default}{
  columns=fullflexible,
  captionpos=b,
  frame=tb,
  breaklines=true,
  rangebeginprefix={(**\ begin\ },
  rangeendprefix={(**\ end\ },
  rangesuffix={\ *)},
  includerangemarker=false,
  basicstyle=\footnotesize\ttfamily,
  identifierstyle={},
  keywordstyle=[1]{\itshape\color{OliveGreen}},
  keywordstyle=[2]{\bfseries\color{Blue}},
  keywordstyle=[3]{\bfseries\color{Red}},
  keywordstyle=[4]{\bfseries\color{Red}},
  keywordstyle=[5]{\bfseries\color{Blue}},
  keywordstyle=[6]{\itshape\color{Blue}},
  keywordstyle=[7]{\color{Blue}},
  keywordstyle=[8]{\color{Red}},
  keywordstyle=[9]{\color{OliveGreen}},
  keywordstyle=[10]{\itshape\color{OliveGreen}},
  keywordstyle=[11]{\itshape\color{OliveGreen}},
  keywordstyle=[12]{\itshape\color{OliveGreen}},
  keywordstyle=[13]{\itshape\color{OliveGreen}},
  keywordstyle=[14]{\itshape\color{OliveGreen}},
  literate={phi}{{$\!\phi\,$}}1
           {phi1}{{$\!\phi_1$}}1
           {phi2}{{$\!\phi_2$}}1
           {phi3}{{$\!\phi_3$}}1
           {phin}{{$\!\phi_n$}}1
}

\lstdefinestyle{easycrypt-pretty}{
    basicstyle=\small\ttfamily,
    literate={:=}{{$\mathrel{\gets}$}}1
              {<=}{{$\mathrel{\leq} \; \;$}}1
              {>=}{{$\mathrel{\geq}$}}1
              {<>}{{$\mathrel{\neq} \; \;$}}1
              {=\$}{{$\stackrel{\$}{\gets}\;$}}1
              {forall}{{$\forall\;$}}1
              {exists}{{$\exists\;$}}1
              {->}{{$\rightarrow\;$}}1
              {<r}{$< \! \! \$ \;$}1
              {<p}{$< \! \! @\;$}1
              {<-}{{$\leftarrow\;$}}1
              {<->}{{$\leftrightarrow\;$}}1
              {<=>}{{$\Leftrightarrow\;$}}1
              {=>}{{$\Rightarrow\;$}}1
              {==>}{{$\Rrightarrow\;$}}1
              {\/\\}{{$\wedge\;$}}1
              {\\\/}{{$\vee\;$}}1
              {.\[}{{[}}1
              {''ora}{{$\mathrel{||}$}}1 
              {'a}{{\color{OliveGreen}$\alpha\,$}}1
              {'b}{{\color{OliveGreen}$\beta\,$}}1
              {'c}{{\color{OliveGreen}$\gamma\,$}}1
              {'t}{{\color{OliveGreen}$\tau\,$}}1
              {'x}{{\color{OliveGreen}$\chi\,$}}1
              {lambda}{{$\lambda\,$}}1
              {sumvd}{{$\sum_{\textrm{v} \in \textrm{d}}$}}1
              {o5}{{$\frac{1}{2}$}}1
              {result}{{$\result$}}1 
              {m0}{{$\textrm{m}_0$}}1
              {m1}{{$\textrm{m}_1$}}1
              {l0}{{$\textrm{l}_0$}}1
}

\begin{document}

\title{Machine-checked ZKP for NP-relations: Formally Verified Security Proofs and Implementations of MPC-in-the-Head}

\author[1]{José Carlos Bacelar Almeida}
\author[1]{Manuel Barbosa}
\author[2]{Karim Eldefrawy}
\author[2]{Stéphane Graham-Lengrand}
\author[1]{Hugo Pacheco}
\author[1,2]{Vitor Pereira}
\affil[1]{University of Porto (FCUP) and INESC TEC}
\affil[2]{SRI International}
\date{}


\maketitle

\vspace{-1.5cm}

\begin{abstract}
MPC-in-the-Head (MitH) is a general framework that allows constructing
efficient Zero Knowledge protocols for general NP-relations from
secure multiparty computation (MPC) protocols. In this paper we give
the first machine-checked implementation of this transformation. 
We begin with an EasyCrypt formalization of MitH that preserves the modular structure of MitH and can be instantiated with arbitrary MPC protocols 
that satisfy standard notions of security, which allows us to leverage 
an existing machine-checked secret-sharing-based MPC protocol development. 
The resulting concrete ZK protocol is proved secure and correct in
\EasyCrypt.
Using a recently developed code extraction mechanism for \EasyCrypt we synthesize a
formally verified implementation of the protocol, which we benchmark
to get an indication of the overhead associated with our formalization
choices and code extraction mechanism.
\end{abstract}


\vfill {\small Distribution Statement ``A'' (Approved for Public Release, Distribution
  Unlimited).\\
  This material is based upon work supported by DARPA under
  Contract No.~HR001120C0086.  Any opinions, findings and conclusions or
  recommendations expressed in this material are those of the author(s) and do
  not necessarily reflect the views of the United States Government or
  DARPA.}

\newpage

\tableofcontents

\section{Introduction}
\label{sec:introduction}

\renewcommand{\P}{\mathcal{P}}
\renewcommand{\V}{\mathcal{V}}

The MPC-in-the-Head (MitH) paradigm was introduced by Ishai, Kushilevitz, Ostrovsky and Sahai~\cite{MitH} as a new foundational bridge between secure multi-party computation (MPC) and zero-knowledge proof (ZK) protocols. 
A ZK protocol for an NP relation $R(x,w)$ can be seen as a two party computation where a prover $\P$ with input $(x,w)$ and a verifier $\V$ with input $x$ jointly compute the boolean function $f(x,w)$ that accepts the proof if and only if $R(x,w)$ holds. 
This means that general feasibility results for two-party secure computation---where no honest majority can be assumed and malicious behavior must be considered---translate into feasibility results for ZK protocols for general relations.

However, the MitH paradigm shows that there exists an {\em efficiency} advantage in considering MPC protocols for $n > 2$ and using a commit-challenge-response transformation to obtain a ZK protocol. Intuitively, $\P$ emulates the execution of a secure computation protocol $\pi$, where $n$ parties compute $f(x,\bar{w})$ and commits to the views of all parties---here $\bar{w}$ denotes a secret sharing of $w$ among the $n$ parties, where any subset of $k$ shares reveals nothing about $w$. 
$\V$ then chooses a (minority) subset of $k$ parties uniformly at random, and $\P$ opens the corresponding views. $\V$ accepts the proof if these views are consistent with an honest execution of $\pi$ by the $k$ parties, where all $k$ parties output $1$. 
The MPC correctness of $\pi$ and binding guaranteed by the commitment scheme suffice to ensure both completeness and a $1/(\frac{n}{k})$ soundness error, where the latter can be reduced by repetition.
Zero knowledge follows from the MPC security of $\pi$ in the presence of an honest majority---the $k$ opened views can be simulated without knowing $w$---as well as the hiding property of the commitment scheme---nothing about $w$ is revealed about the remaining $n-k$ parties' views.

The efficiency gain of MitH stems from two important observations: 
1) that $\pi$ only needs to satisfy a weak notion of security that allows for extremely efficient instantiations and 
2) that the round complexity of $\pi$ has no impact in the final protocol, since $\pi$ is evaluated ``in-the-head''.
A series of follow-up works~\cite{KatzBoo,ZKBoo,ZKBoopp,baum2020concretely,Ligero,Ligeropp} demonstrated the efficiency and flexibility of protocol families inspired in the MitH core ideas, by exploring adaptations of this principle to specific MPC protocols and different variants of ZK protocols. One notable takeaway of these works is that MitH offers a high degree of flexibility, as it allows for instantiations that can efficiently handle relations that are more naturally expressed as both arithmetic and boolean circuits.

In this work we explore the elegant simplicity and modularity of the MPC-in-the-Head paradigm to obtain machine-checked security proofs and formally-verified implementations for ZK protocols supporting general relations.
To the best of our knowledge this is the first end-to-end machine-checked development of ZK-protocols for all relations in NP.
We focused on the MitH variant that can be instantiated with passively secure secret-sharing-based MPC protocols that tolerate two corrupt (i.e., semi-honest) parties. 
This allows us to build on an existing development that already provides a suitable instantiation for the underlying MPC protocol, secret sharing and commitment schemes.
On the other hand, this opens the way for future work formalizing optimizations of the construction that follow along the same path~\cite{ZKBoo,ZKBoopp}.%
\footnote{We note that different performance trade-offs can be explored when the underlying MPC protocol offers security against active adversaries; we do not consider these here because the most competitive constructions that follow this path~\cite{Ligero,Ligeropp} follow a more intricate modular construction to convert the underlying MPC protocol to the resulting ZK protocol based on probabilistically checkable proofs (PCP). Formally verifying these protocol families is also an interesting direction for future work.}

In more detail, our contributions are as follows:
\begin{itemize}
\item We formally specify and verify the foundational results of Ishai et al~\cite{MitH} and create a generic framework that permits the modular specification and formal verification of secret-sharing-based MPC protocols for general circuits, which can then be plugged into a formally-verified MitH generic transformation to produce provably secure ZK protocols for general relations.
We give a complete formalization of the general construction and its potential instantiations in \EasyCrypt. 

\item For the MPC protocol, we reuse the formally-verified implementation of the BGW protocol given in~\cite{priormpc} instantiated to $n=5$ parties.
For the commitment scheme, we have two instantiations; the first reuses the Pedersen commitment scheme formalized in~\cite{priormpc}, and the second relies on a more efficient PRF-based construction that we instantiate with HMAC~\cite{hmaccommitment}.
The machine-checked security proofs for the MPC protocol and commitment scheme directly transpose those found in the literature, which highlights the fact that our framework can be instantiated with standard components.

\item For the generic construction, we formalize the security proofs for completeness, soundness and zero-knowledge at two levels: 1) we give a proof of perfect completeness, as well as concrete bounds for the soundness error and for the zero-knowledge simulation strategy when the malicious party (prover and verifier, respectively) is executed once;  2) we then complement these results with formal machine-checked proofs of the meta arguments that require repetition using the \EasyCrypt operator logic. 
In this way we can express in the \EasyCrypt logic the repeated independent executions of the same adversary. 
We chose {\em not} to idealize the underlying commitment scheme and formalize the proof as in~\cite[Theorem 3.2]{MitH}, which makes the proof verification more challenging, but yields a result that can be instantiated with standard commitment schemes.

\item We use the automatic code extraction mechanism for \EasyCrypt proposed in \cite{priormpc} to automatically extract a verified implementation of the full Zero-Knowledge protocol produced by the MitH transformation and give preliminary benchmarking results. Our results are promising and open the way for future work where verified implementations of MitH protocols essentially match the performance of non-verified ones.

\end{itemize}



\paragraph{Access to development.} Our EasyCrypt formalization of the MitH protocol, as well as its extracted code, can be found at \href{https://github.com/SRI-CSL/high-assurance-crypto/tree/main/high-assurance-zk}{https://github.com/SRI-CSL/high-assurance-crypto/tree/main/high-assurance-zk}.

\section{Related work}
\label{sec:related-work}
%
We give an overview of the most relevant work in the field of computer-aided cryptography. This is an active area of research that aims to address the challenges of achieving formally robust, machine-checked, and practically efficient cryptographic protocols, which gave rise to several software developments and formal verification tools~\cite{SP:BarbosaBBBCLP21}.
At the design level, tools allow describing the high level specification of the protocol and stating its security properties while managing the complexity of the security proofs via a modular design of the various components.
At the implementation level, tools help to establish the functional correctness of (often aggressively optimized) implementations w.r.t. the specification. In our approach, the specification considers the security of abstract cryptographic primitives and representative concrete instantiations; a correct-by-construction implementation for each concrete instantiation is then automatically extracted from the specification. This highlights the modularity and generality of the formalization and grants more flexibility for exploring different instantiations and extracting the respective reference implementations, in contrast to devising and verifying a highly optimized for a particular instantiation.

Outside the realm of computer aided cryptography, cryptographers have devised methodological frameworks for performing security proofs, such as code-based game playing~\cite{gameplaying}. The motivation for these frameworks is to systematize and the decompose security arguments to minimize the likelihood of errors, even in ``pen-and-paper'' proofs.
The \EasyCrypt tool supports a general-purpose relational program logic that is close to code-based game playing techniques. Therefore, \EasyCrypt proofs tend to follow the structure of pen-and-paper proofs.

%
Zero Knowledge protocols are a fast-moving area within cryptography, and many protocols exist both for specific proof goals and for settings that require a flexible  solution that can be used for any relation.
Only a very small part of this field has been studied from the perspective of computer aided cryptography.
The work in~\cite{sigmaCSF} was the first to formalize a special class of $\varSigma^{\phi}$-protocols in \CertiCrypt, a predecessor of \EasyCrypt implemented as a \Coq library, and to prove the security of general $\textit{and}$ and $\textit{or}$ composability theorems for $\varSigma^{\phi}$-protocols. The more recent work from~\cite{ZKCryptHOL} restates many of these results for $\varSigma$-protocols in \CryptHOL. It additionally formalizes abstract and concrete commitment scheme primitives and proves a construction of commitment schemes from $\varSigma$-protocols.

The most significant machine checked endeavor for ZK is the work in~\cite{fullzk}, that developed a full-stack verified framework for developing ZK proofs. The framework encompasses a non-verified optimizing ZK compiler that translates high-level ZK proof goals to C or Java implementations, and a verified compiler that generates a reference implementation. The machine checked effort lies in proving that, for any goal, the reference implementation satisfies the ZK properties and that the optimized implementation has the same observable behavior as the reference implementation. The core of the verified compiler builds on top of the results from~\cite{sigmaCSF}, extended with $\textit{and}$ compositions of $\varSigma^{\texttt{GSP}}$-protocols, and generates \CertiCrypt proof scripts for automatically proving the equivalence of the two implementations.

%
There is nowadays a vast body of MPC protocols and frameworks, some of which have been formally verified using machine-checked tools.
CircGen~\cite{almeida2017fast} is a verified compiler translates C programs into boolean circuits by extending the CompCert C compiler with an additional backend translation to Boolean circuits. This back-end can then be used to feed to a \EasyCrypt machine-checked implementation of Yao's 2-party secure function evaluation protocol.
The work in~\cite{haagh2018computer} formalizes in \EasyCrypt the $n$-party MPC protocol due to Maurer~\cite{maurer2006secure} for the actively secure case.
The work in~\cite{priormpc} provides verified implementations of pro-actively secure MPC, including an \EasyCrypt formalization of the BGW~\cite{ben2019completeness} MPC protocol for passive and static active adversaries that we adapt and reuse in this paper.

\section{Preliminaries}
\label{sec:preliminaries}

We give in this section the formal cryptographic definitions that we use
in our formalization, which are all standard. 
We follow~\cite{MitH} closely.
We also give a short overview
of the \EasyCrypt features required to understand the presentation of our
development.

\subsection{Basic Primitives}
\label{sec:basicprimitives}

\newcommand{\W}{\mathcal{W}}
\newcommand{\Shr}{\mathsf{Shr}}
\newcommand{\USh}{\mathsf{USh}}
\newcommand{\ShS}{\mathsf{ShS}}

The MitH construction requires a very simple form of secret sharing over the inputs of the MPC protocol, although the underlying MPC protocol may internally use additional properties we will not mention here.

We also require standard commitment schemes. 
Note that, as the MPC protocols and secret sharing protocols we use to instantiate
the MitH construction are information theoretically secure, the security of the
commitment scheme determines the class of adversaries over the security of the
MitH construction. In particular, the standard soundness property
only holds if the commitment is statistically hiding; if the commitment scheme
is only computationally binding, then we obtain a ZK argument.

\begin{definition}[\bf Secret Sharing Scheme]
For a secret space $\W$, which we will take to be the same as a share space,
a secret sharing scheme for $n$ parties is defined by two ppt algorithms:
\begin{itemize}
\item The probabilistic sharing algorithm $\Shr(w)$ takes a secret $w \in \W$ 
and returns a sharing $\bar{w} \in \W^n$.
\item The deterministic reconstruction algorithm $\USh(\bar{w})$ takes
a sharing $\bar{w} \in \W^n$ and returns a secret $w \in \W$.
\end{itemize}
A secret sharing scheme should satisfy the following properties:
\begin{itemize}
    \item{\sc Correctness}: For all secrets $w \in \W$ and
    all sharings $\bar{w} \in [\Shr(w)]$, we have that 
    $w = \USh(\bar{w})$. Moreover, $\USh$ is a total function
    over $\W^n$, which means that all sharings determine a
    unique secret.
    \item{\sc $t$-Privacy} Any subset of $t$ shares, for $0 \le t < n$ 
    can be simulated efficiently by sampling from a fixed public distribution. 
    Formally, we require that there exist a simulated share distribution 
    $\ShS$ such that, for any subset $S \subseteq [n]$ such that $|S| = t$ 
    and all $w \in \W$:
\[
\{ \, [\bar{w}]_{i \in S} \, : \, \bar{w} \sample \Shr(w) \, \} 
\ \equiv  \ 
\{ \, \hat{w} \,  : \, \hat{w} \sample \ShS \, \} 
\]
\end{itemize}
\end{definition}

\newcommand{\MS}{\mathcal{M}}
\newcommand{\CS}{\mathcal{C}}
\newcommand{\KS}{\mathcal{K}}
\newcommand{\A}{\mathcal{A}}
\renewcommand{\Com}{\mathsf{Com}}
\renewcommand{\Ver}{\mathsf{Ver}}

\begin{definition}[\bf Commitment Scheme]
For a message space $\MS$, a commitment space $\CS$ and an opening space
$\KS$, a commitment scheme is defined by two ppt algorithms:
\begin{itemize}
\item The probabilistic commitment algorithm $\Com(m)$ takes a secret $m \in \MS$ 
and returns a pair $(c,k)$ where $c \in \CS$ is a commitment and $k \in \KS$
is an opening.
\item The deterministic verification algorithm $\Ver(m,c,k)$ takes
a message $m \in \MS$, a commitment $c \in \CS$, and an opening $k \in \KS$
and it returns $1$ or $0$ indicating success or failure, respectively.
\end{itemize}
A commitment scheme should satisfy the following properties:
\begin{itemize}
    \item{\sc Correctness:} For all messages $m \in \MS$, all 
    commitments $c \in \CS$ and openings $k \in \KS$, such that 
    $(c,k) \in [\Com(m)]$, we have that $\Ver(m,c,k)=1$.
    \item{\sc Binding:}
    The commitment scheme is computationally (resp. statistically) binding if the probability that the following game returns $1$ is bounded by a 
    small $\epsilon_b$ when $\A$ is a ppt (resp. a potentially unbounded) algorithm:
     \begin{itemize}
       \item the binding game first runs an adversary $\A$ who outputs a tuple $(m,k,m',k',c)$;
       \item the game then runs $\Ver(m,c,k)$ and $\Ver(m',c,k')$.
       \item the game returns $1$ if and only if both verifications are successful and $m \ne m'$.
   \end{itemize}
    \item{\sc Hiding:} The commitment scheme is computationally (resp. statistically) hiding if the
    left or right indistinguishability advantage of an adversary $\A$ in the following game is bounded by a small $\epsilon_h$ bias, when $\A$ is a ppt (resp. a potentially unbounded) algorithm:
    \begin{itemize}
     \item the hiding game first samples a random coin $b$ and runs adversary $\A$ that chooses two messages $m_0, m_1 \in \MS$;
     \item the game then computes $(c,k) \sample \Com(m_b)$, which it provides to $\A$;
     \item eventually $\A$ terminates outputting a guess $b'$ and the game outputs $1$ if $b = b'$ and $0$ otherwise.
 \end{itemize}
  The advantage of  $\A$ is defined  in  the standard way as $|p - 1/2| \le \epsilon_h$, where $p$ is the probability that the hiding game returns $1$.
\end{itemize}
\end{definition}

\subsection{Zero-Knowledge}
\label{sec:zero-knowl-proofs}

A NP-relation $R(x,w)$ is an efficiently decidable and polynomially bounded binary relation, which we see as a boolean function. 
This implies that, for all $(x,w)$, if $R(x,w)=1$, then $|w| \le p(|x|)$ for some fixed polynomial $p$.

A ZK protocol for a NP-relation $R(x, w)$ is defined by two ppt interactive algorithms, a prover $\P$ and a verifier $\V$. 
The prover takes a NP statement $x$ and a corresponding witness $w$; the verifier is only given the statement $x$. 
The prover and the verifier interact---in this paper we consider only three-pass commit-challenge-response protocols---until eventually the verifier outputs $1$ or $0$ indicating success or failure, respectively.
The view of $\V$ is defined as its input $x$, its coin tosses and all the messages that it receives.%
\footnote{Throughout the paper we will often see interactive protocols as a sequence of calls to {\em next-message} functions: these functions take the coin tosses of the corresponding party and all received messages and they deterministically define the next message transmitted by the party. In particular, next message functions take partial views and define the next message  to be sent. A complete view of a party defines its output, which we see as a particular case of the next-message function output.}

\renewcommand{\S}{\mathcal{S}}

\begin{definition}[\bf Zero-knowledge proof]
A protocol $(\P, \V)$ is a zero-knowledge proof protocol for the relation $R$ if it satisfies the following requirements:
\begin{itemize}
\item {\sc Completeness}: In an honest execution, if $R(x, w) = 1$, then the verifier  accepts with probability $1$.
\item{\sc Soundness}: For every malicious and computationally unbounded prover $\P^*$, there is a negligible function $\epsilon(\cdot)$ such that, if $R(x, w) = 0$ for all $w \in \{0,1\}^{p(|x|)}$, then $\P^*$ can make $\V$ accept with probability at most $\epsilon(|x|)$.
\item {\sc Zero-Knowledge}: For any malicious ppt verifier $\V^*$, there exists a ppt simulator $\S^*$, such that the view of $\V^*$ when interacting with $\P$ on inputs $(x, w)$ for which $R(x, w) = 1$, is computationally indistinguishable from the output of $\S$ on input $x$.%
\footnote{Throughout the paper,  when we say that two distributions are computationally (resp. statistically) indistinguishable we mean that, for all ppt (resp. potentially unbounded) distinguishers returning a bit, the probability that the distinguisher returns $1$ when fed with a value sampled from either of the distributions changes by a small quantity $\epsilon$. For perfect indistinguishability, $\epsilon$ is $0$ and we consider unbounded distinguishers.}
\end{itemize}
\end{definition}

\noindent
We will also consider zero-knowledge protocols that have a constant (non-negligible) soundness error $\epsilon$, in which cases the soundness error will be specified.

\subsection{Secure Multiparty Computation}
\label{sec:secure-mult-comp}

The MitH paradigm builds on secure function evaluation protocols that assume synchronous communication over secure point-to-point channels. 

Let $n$ be the number of parties, which will be denoted by $P_1,\ldots,P_n$. 
All players share a public input $x$, and each player $P_i$ holds a local private input $w_i$. We consider protocols that can securely compute a $n$-input function $f$ that maps the inputs $((x,w_1),\ldots,(x,w_n))$ to a $n$-tuple of boolean outputs.

A protocol $\Pi$ is specified via its next message function. 
That is, $\Pi(i, x, w_i , r_i , (m_1 , \ldots , m_j ))$ returns the set of $n$ messages sent by $P_i$ in round $j + 1$, given the public input $x$, its local input $w_i$, its random input $r_i$, and the messages $m_1, \ldots , m_j$ it received in the first $j$ rounds. 
The output of the next message function $\Pi$ may also indicate that the protocol should terminate, in which case $\Pi$ returns the local output of $P_i$.

\newcommand{\view}{V}

The view of $P_i$, denoted by $\view_i$, includes $w_i$, $r_i$ and the messages received by $P_i$ during the execution of $\Pi$. 
Note that $\Pi$ and $\view_i$ fully define the set of messages sent by $P_i$ and also its output.
The following notion of consistent views and relation between local and global consistency are important for the MitH transformation.

\begin{definition}[\bf Consistent views] 
A pair of views $\view_i, \view_j$ are consistent (with respect to the protocol $\Pi$ and some public input $x$) if the outgoing messages implicit in $\view_i$ are identical to the incoming messages reported in $\view_j$ and vice versa.
\end{definition}

\begin{lemma}[\bf Local vs. global consistency~\cite{MitH}] 
\label{th:consistency}
Let $\Pi$ be a $n$-party protocol as above and $x$ be a public input. 
Let $\view_1,\ldots,\view_n)$ be an $n$-tuple of (possibly incorrect) views. 
Then all pairs of views $\view_i,\view_j$ are consistent with respect to $\Pi$ and $x$ if and only if there exists an honest execution of $\Pi$ with public input $x$ (and some choice of private inputs $w_i$ and random inputs $r_i$) in which $\view_i$ is the view of $P_i$ for every $1 \le i \le n$.
\end{lemma}

\noindent
We consider security of protocols in the semi-honest model, where we define correctness and privacy as follows.

\begin{definition}[\bf Correctness]
Protocol $\Pi$ securely computes function $f((x,w_1),\ldots,(x,w_n))$ with perfect correctness if, for all inputs $x, w_1, \ldots, w_n$, 
the probability that the output of some player is different from the output of $f$ is 0, where the probability is over the independent choices of the random inputs $r_1,\ldots , r_n$.
\end{definition}

\begin{definition}[\bf $t$-Privacy]
We say that $\pi$ securely computes $f$ with perfect $t$-privacy,  for $1 \le t \le n$, if there is a ppt simulator $\S$ such that, for any inputs $x, w_1, \ldots , w_n$ and every set of corrupted parties $T \subseteq [n]$, where $|T| \le t$, the joint view $(\view_{i_1},\ldots,\view_{i_k})$ of the parties in $T = \{i_1,\ldots,i_k\}$ is computationally indistinguishable from $\S(T, x, [w_i]_{i\in T}, [f(x, w_1, \ldots , w_n)_i]_{i \in T})$. 
\end{definition}

\paragraph{Remark.} 
For the concrete protocols we consider to instantiate the MitH transformation, $\bar{w}=(w_1,\ldots,w_n)$ will be a secret sharing of the witness that is input to the ZK protocol. 
Furthermore, function $f$ will be the boolean function that computes $R(x,w)$, where $w$ is the secret that results from the reconstruction of $\bar{w}$, and outputs the result to all parties.

\subsection{Background on \EasyCrypt}
\label{sec:form-verif-code}

\EasyCrypt is an interactive proof assistant tailored for the
verification of the security of cryptographic constructions in the
computational model. \EasyCrypt adopts the code-based approach, in
which primitives, security goals and hardness assumptions are
expressed as probabilistic programs. The \EasyCrypt proof system
incorporates an ambient logic\---for first-order
and higher-order logic reasoning through a set of interactive proof
tactics\---and a probabilistic relational Hoare logic\---%
for reasoning about programs written in an imperative
C-like language, as well as establishing equality results between two
program executions. Moreover, \EasyCrypt uses formal tools from program
verification, particularly SMT provers, that can be used in
cooperation with the ambient logic to discard proof goals.

\EasyCrypt supports a high degree of modular reasoning via its theory
system. In short, a theory is a collection of types and operators that
can later be refined or extended with more data
structures. Alternatively, it can be seen as an abstract class that fixes
the interface that \textit{objects} (i.e., concrete instantiations of
that theory) must follow. \EasyCrypt also provides a cloning mechanism
that allows theories to refine the types and operators of more abstract theories, or to combine the elements of existing ones.

Another relevant \EasyCrypt specificity that we widely use in our
formalization is the ability to establish a \textit{module type} (i.e., an abstract interface), that
defines the signature of procedures that an \EasyCrypt module
implementing that type are expected to follow.
Module types are very useful because the \EasyCrypt proof system permits
the universal quantification over all possible implementations of the
module type. Moreover, they can also be used as parameters to
other modules, in order to define generic constructions based on
abstract primitives.

The \EasyCrypt development, as well as the SMT solvers used in cooperation with the \EasyCrypt proof system are part of our trusted computing base
(TCB).  
The \EasyCrypt code extraction tool has been recently
developed in \cite{priormpc} is also part of the TCB. 
This tool allows to extract \OCaml
executable code from an \EasyCrypt proof script where code is defined
via functional operators.

\section{Modular machine-checked proof of MPC-in-the-Head}
\label{sec:modul-mach-check}

\begin{figure}[!ht]
  \centering
  \includegraphics[width=0.8\textwidth]{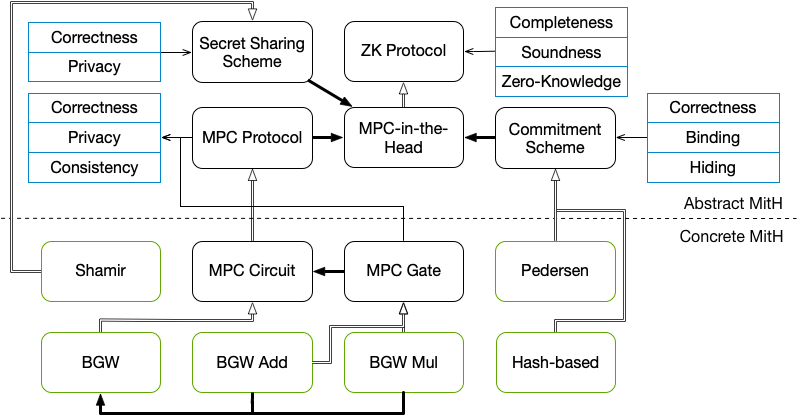}
  \caption{
    Bird's eye view of the relation between
    different parts of our formalization.
    The horizontal dotted line separates abstract MitH components (above) from our proof-of-concept concrete instantiations (below).
    Rounded rectangles represent cryptographic
    constructions. Rounded green rectangles represent
    concrete instantiations of the respective cryptographic
    primitives. Blue rectangles depict security
    definitions. Arrows $\rightarrow$ represent security definitions
    applied to the connected cryptographic constructions.
    Double arrows $\Rightarrow$ entail that a component instantiates a more abstract component, and in particular provides concrete proofs for its security definitions.
    Finally, if two boxes are connected via a
    bold arrow \ding{225},
    then one is a sub-component of the
    other.}
  \label{fig:mith-formal-proof-arch}
\end{figure}

Our \EasyCrypt development defines an abstract and modular
infrastructure that follows the general MitH transformation, 
and can be instantiated with different concrete components.
%
Figure
\ref{fig:mith-formal-proof-arch} depicts the relation between
these different components in our formalization, and includes
abstractions for the syntax and security of ZK protocols, 
secret sharing schemes, MPC protocols and commitment schemes,
as introduced in Section~\ref{sec:preliminaries}.
Our syntax for MPC protocols fixes an abstract notion of
a circuit, which is used to formalize the semantics of
evaluating a functionality.
On top of these abstractions we specify the MitH transformation
and prove that it can be used to construct ZK protocols for 
NP-relations expressed as circuit satisfiability.
Finally, we formalized concrete instantiations of each component, 
which are then put together to obtain fully concrete instantiations 
of ZK protocols via the general MitH transformation.
Our concrete instantiations can handle proof goals expressed
using {\em arithmetic} circuits, as we rely on the 
BGW~\cite{ben1988completeness} protocol under the hood.
However, our framework is general enough to allow for
different instantiations, including MPC protocols that
rely on different circuit representations.
In what follows, we will provide a more detailed view of our formalization, resorting directly to snippets of \EasyCrypt code.

\subsection{ZK protocols}

\paragraph{Syntax.}
The \ec{ZKProtocol} theory fixes the syntax of ZK protocols in
EasyCrypt, i.e., the types and operators that
must be defined by a concrete ZK protocol.
We fix the commit-challenge-response three pass
protocol structure, since this is all we need for the MitH
transformation.

  \begin{lstlisting}[mathescape,language=easycrypt,xleftmargin=0pt,xrightmargin=0pt,style=easycrypt-pretty,basicstyle=\small]
theory ZKProtocol.
  type witness_t.
  type statement_t.

  op relation : witness_t -> statement_t -> bool.
  op language(x : statement_t) = exists w, relation w x = true.

  type prover_input_t = witness_t * statement_t.
  type verifier_input_t = statement_t.

  type prover_state_t.
  type verifier_state_t.

  type prover_rand_t.
  type verifier_rand_t.

  type prover_output_t = unit.
  type verifier_output_t = bool.

  type commitment_t. 
  type challenge_t. 
  type response_t.

  op commit : prover_rand_t -> prover_input_t -> prover_state_t * commitment_t.
  op challenge  : verifier_rand_t -> verifier_input_t -> commitment_t -> verifier_state_t * challenge_t.
  op response   : prover_state_t -> challenge_t -> response_t.
  op check : verifier_state_t -> response_t -> bool.

  type trace_t = commitment_t * challenge_t * response_t.

  op protocol (r : prover_rand_t * verifier_rand_t) (x : prover_input_t * verifier_input_t) : trace_t * (prover_output_t * verifier_output_t) = 
    let (r_p, r_v) = r in let (x_p, x_v) = x in
    let (st_p,c) = commit r_p x_p in
    let (st_v,ch) = challenge r_v x_v c in
    let r = response st_p ch in
    let b = check st_v r in ((c, ch, r), ((),b)).
end ZKProtocol.
\end{lstlisting}

Types that are undefined at this level must be specified by
each protocol. This is the case, for example, for the type
of witnesses and statements, but not for the outputs of the
prover and verifier, which are hardwired in the syntax to
be the singleton type and a boolean value, respectively.
Each protocol is associated with a relation, 
which at this level is modeled as an abstract
boolean function. 
Finally, the theory also defines what it means to
honestly execute the protocol via the \ec{protocol}
operator.

\paragraph{Completeness.}
The completeness game is parameterized by a randomness generator
algorithm \ec{R}. Each protocol needs to define what it means to sample
the random coins of the prover and verifier using this mechanism.%
\footnote{This pattern will appear in the remaining games, so we will omit
further descriptions of randomness generation modules.}
The \EasyCrypt formalization is a direct translation of the
completeness requirement given in Section~\ref{sec:preliminaries}.

\begin{lstlisting}[mathescape,language=easycrypt,xleftmargin=0pt,xrightmargin=0pt,style=easycrypt-pretty,basicstyle=\small]
module Completeness(R : Rand_t) = {
  proc main(w : witness_t, x : statement_t) : bool = {
    var r_p, r_v, tr,y;
    (r_p, r_v) <@ R.gen();
    (tr,y) <- protocol (r_p, r_v) ((w,x), x);
    return (snd y);
  }
}.
\end{lstlisting}

\noindent
A completeness claim in \EasyCrypt can be written as:
\begin{lstlisting}[mathescape,language=easycrypt,xleftmargin=0pt,xrightmargin=0pt,style=easycrypt-pretty,basicstyle=\small]
forall w x, relation w x => Pr [ Compleneness(R).main(w,x) : res ] = 1%r.
\end{lstlisting}

\paragraph{Soundness.}
For the soundness definition, we need to quantify over potentially
malicious provers. In \EasyCrypt this is done by defining a
module type, i.e., the interface that the adversary exposes
Module type \ec{MProver_t} specifies this interface.

\begin{lstlisting}[mathescape,language=easycrypt,xleftmargin=0pt,xrightmargin=0pt,style=easycrypt-pretty,basicstyle=\small]
module type MProver_t = {
  proc commitment (x: statement_t) : commitment_t
  proc response(x : statement_t, c : commitment_t, ch : challenge_t) : response_t
}.
\end{lstlisting}

The soundness property can now be expressed as a game which is
parameterized by an attacker of this type. This allows us to quantify
universally over malicious provers.

\begin{lstlisting}[mathescape,language=easycrypt,xleftmargin=0pt,xrightmargin=0pt,style=easycrypt-pretty,basicstyle=\small]
module Soundness(RV : RandV_t, MP : MProver_t) = {
  proc main(x : statement_t) : bool = {
    var r_v, c, ch,st_r,  resp;
    r_v <@ RV.gen();
    c <@ MP.commitment(x);
    (st_r, ch) <- challenge r_v x c;
    resp <@ MP.response(x, c, ch);
    return (!language x /\ check st_r resp);
  }
}.
\end{lstlisting}

\noindent
A soundness claim in \EasyCrypt can be written as:
\begin{lstlisting}[mathescape,language=easycrypt,xleftmargin=0pt,xrightmargin=0pt,style=easycrypt-pretty,basicstyle=\small]
forall x MP, Pr [ Soundness(RV,MP).main(x) : res ] <= epsilon.
\end{lstlisting}

\paragraph{Zero Knowledge.}
We formalize the zero knowledge property using the following
experiment in which a (malicious) verifier interacts with an
honest prover or with the simulator. We will see below how the evaluator
module \ec{E} is defined for either case.

\begin{lstlisting}[mathescape,language=easycrypt,xleftmargin=0pt,xrightmargin=0pt,style=easycrypt-pretty,basicstyle=\small]
module ZKGame (D : Distinguisher_t) (RP : RandP_t) (E : Evaluator_t) (MV : MVerifier_t) = {
  proc main(w : witness_t, x : statement_t) : bool option = {
    var rp, b, check, tr;
    rp <@ RP.gen();
    (check,tr) <@ E.eval(w,x,rp);
    b <@ D.guess(w,x,check,tr);
    return b;
  }
}.
\end{lstlisting}

For both cases we need to define the type of malicious verifier modules
\ec{MVerifier_t}, which we omit for brevity.
To define the evaluator module for the simulated view, we
first specify a type for the simulator which acts like a
prover.

\begin{lstlisting}[mathescape,language=easycrypt,xleftmargin=0pt,xrightmargin=0pt,style=easycrypt-pretty,basicstyle=\small]
module type Simulator_t = {
  proc commitment(x : statement_t) : commitment_t option
  proc response(x : statement_t, ch : challenge_t) : response_t option
}.
\end{lstlisting}

Note however that, unlike a prover, the simulator response may signal an abort (\ec{option} means that the
procedure may sometimes fail to produce a result), often triggered by
a mismatch between the received challenge and some initial guess.
Of course, we shall bound the probability of such simulation failures
as it will have an immediate impact on the Zero Knowledge property.

The evaluator module that animates the
interaction between the simulator and the malicious verifier is defined as follows.

\begin{lstlisting}[mathescape,language=easycrypt,xleftmargin=0pt,xrightmargin=0pt,style=easycrypt-pretty,basicstyle=\small]
module (IdealEvaluator(S : Simulator_t) : Evaluator_t)  (MV : MVerifier_t) = {
  proc eval(w : witness_t, x : statement_t, rp : prover_rand_t) : bool option = {
      var c, ch, oresp, resp, acceptance, ret, oc;
      ret <- None;
      oc <@ S.gen_commitment(x);
      if (oc <> None) {
        c <- oget oc;
        ch <@ MV.gen_challenge(x, c);
        oresp <@ S.gen_response(x, ch);
        if (oresp <> None) {
          resp <- oget oresp;
          acceptance <- check x c ch resp;
          ret <- Some (acceptance, (oget oc,ch, oget oresp));
        }
      }
      return ret;
  }
}.
\end{lstlisting}

For a real protocol execution, the \ec{ZKGame} shall be parametrized 
by the following evaluator module that encodes the normal protocol
execution, and never fails.

\begin{lstlisting}[mathescape,language=easycrypt,xleftmargin=0pt,xrightmargin=0pt,style=easycrypt-pretty,basicstyle=\small]
module (RealEvaluator : Evaluator_t) (MV : MVerifier_t) = {
  proc eval(w : witness_t, x : statement_t, rp : prover_rand_t) = {
    var st_p, c, ch, resp,b;
    (st_p, c) <- commit rp (w,x);
    ch <@ MV.gen_challenge(x, c);
    resp <- response st_p ch;
    acceptance <- check x c ch resp;
    return Some (acceptance, (c,ch,resp));
  }
}.
\end{lstlisting}

A zero-knowledge claim, for a concrete simulator
\ec{S} in \EasyCrypt can be written as:

\begin{lstlisting}[mathescape,language=easycrypt,xleftmargin=0pt,xrightmargin=0pt,style=easycrypt-pretty,basicstyle=\small]
forall w x (D <: Distinguisher_t) (MV <: MVerifier_t), relation w x => 
    | ZKGame(RP,RealEvaluator,MV).main(w,x) : res ] - 
       ZKGame(RP,IdealEvaluator(S),MV)).main(w,x) : res ] | <= epsilon
\end{lstlisting}

\subsection{MPC protocols}
\label{sec:mpccomponent}

Our formalization of MPC protocols shares many similarities with the
\ec{ZKProtocol}.
The main difference, in addition to considering an arbitrary number 
of parties, is that we need to consider protocols that are parametric
on an abstract type for circuits (i.e., a way to represent $n$-input to
$n$-output computations)
and an abstract operator that defines what it means to evaluate an
arbitrary circuit.
The \ec{Protocol} theory is similar to that used in~\cite{priormpc}.

\begin{lstlisting}[mathescape,language=easycrypt,xleftmargin=0pt,xrightmargin=0pt,style=easycrypt-pretty,basicstyle=\small]
theory MPCProtocol.
  const n : {int | 2 <= n} as n_pos.
  const t : {int | 0 <= t < n} as t_pos.
  type pid_t.
  type pid_set = pid_t list.
  type circuit_t.
  
  type pinput_t. 
  type sinput_t.
  type input_t = pinput_t * sinput_t.

  type output_t.
  op f : circuit_t -> (pid_t * input_t) list -> (pid_t * output_t) list

  type rand_t.
  
  type msgs_t.
  type in_messages_t.
  type out_messages_t.
  op get_messages_from : pid_t -> in_messages_t -> msgs_t.
  op get_messages_to : pid_t -> out_messages_t -> msgs_t.

  type trace_t = in_messages_t.
  
  type view_t = input_t * rand_t * in_messages_t.
  type views_t = (pid_t * view_t) list.

  op protocol : circuit_t -> (pid_t * rand_t) list -> (pid_t * input_t) list ->
                                             (pid_t * trace_t) list * (pid_t * output_t) list.

  op out_messages : circuit_t -> pid_t -> view_t -> out_messages_t.
  op local_output : circuit_t -> pid_t -> view_t -> output_t.
end MPCProtocol.
\end{lstlisting}

Party inputs \ec{input_t} can be defined as having a public
and a secret part, and these should not be interpreted as 
a single input wire to the circuit. Indeed, it is the 
circuit evaluation operator \ec{f} that defines the semantics
of evaluating a circuit on given inputs.
The operators that define the behavior of parties are 
\ec{out\_messages} and \ec{local_output}, which match
the next-message function approach introduced in 
Section~\ref{sec:preliminaries}.
Finally, the \ec{protocol} operator is used to define the
global protocol evaluation, which allows for some flexibility
in defining the message scheduling.

Given these definitions we can capture the notion of pairwise 
view consistency as the following \EasyCrypt predicate,
where \ec{pinput} is a simple operator that extracts the
public input from the view; the \ec{get_messages\_from},
\ec{get\_messages\_to} and \ec{in\_messages} operators
simply project views and lists of output messages to the
relevant party identifiers for comparison.

\begin{lstlisting}[mathescape,language=easycrypt,xleftmargin=0pt,xrightmargin=0pt,style=easycrypt-pretty,basicstyle=\small]
op consistent_views (c : circuit_t) (xp : pinput_t) (vi vj : view_t) (i j : pid_t) =
  let (xi, ri, ti) = vi in
  let (xj, rj, tj) = vj in
  let outj = out_messages c j vj in
  valid_rand c ri /\ valid_rand c rj /\
  xi.`1 = xp /\ xj.`1 = xp /\ valid_circuit_trace c ti /\ valid_circuit_trace c tj /\ 
  get_messages_from j ti = get_messages_to i tj /\
  get_messages_from i tj = get_messages_to j ti.
\end{lstlisting}

The correctness of MPC protocols is formalized analogously to 
completeness for ZK protocols, so we omit some details.
The $t$-privacy property is formalized using the same approach
used for the zero-knowledge property, the main difference
being that the simulator must now construct $t$ views to be fed to a distinguisher. We give here only the details
of the real view evaluator and the ideal view evaluator for
a fixed simulator \ec{S} with given module type.
In the real-world, the \ec{extract\_views} operator returns
the views of the corrupt parties.
In the ideal world, the \ec{extract\_inputs} operator extracts
the full inputs of corrupt parties. 
This is sufficient because we assume throughout that the
public inputs to all parties are identical, consistently
with Section~\ref{sec:preliminaries}.

\begin{lstlisting}[mathescape,language=easycrypt,xleftmargin=0pt,xrightmargin=0pt,style=easycrypt-pretty,basicstyle=\small]
module type Simulator_t = {
  proc simulate(c : circuit_t, xs : (pid_t * input_t) list, rs : (pid_t * rand_t) list, cr : pid_t list, ys : ProtocolFunctionality.output_t) : views_t
}.

module RealEvaluator = {
  proc eval(c : circuit_t, xs : (pid_t * input_t) list, cr : pid_t list, rs : (pid_t * rand_t) list) = {
     var tr,y,vsc;
    (tr, y) <- protocol c xs rs;
    vsc <- extract_views cr tr;
    return vsc;
  }
}.

module IdealEvaluator (S : Simulator_t) = {
  proc eval(c : circuit_t, xs : (pid_t * input_t) list, cr : pid_t list, rs : (pid_t * rand_t) list) = {
    var ins,vsc;
    ins <- extract_inputs cr xs;
    y <- f c xs;
    vsc <- S. simulate(c, xsc, rs, cr, y);
    return vsc;
  }
}.
\end{lstlisting}

A $t$-privacy claim for a $n$-party protocol 
can be written in \EasyCrypt as follows.
\begin{lstlisting}[mathescape,language=easycrypt,xleftmargin=0pt,xrightmargin=0pt,style=easycrypt-pretty,basicstyle=\small]
forall c x aux, 
    | Pr [ PrivGame(D,R,RealEvaluator).main(c,x,aux) : res ] - 
        Pr [ PrivGame(D,R,IdealEvaluator(S)).main(c,x,aux) : res ] | <= epsilon
\end{lstlisting}

\subsection{Basic Primitives}

\paragraph{Secret Sharing.}
The \EasyCrypt formalization of a secret sharing scheme is straightforward.
The syntax is defined as follows.

\begin{lstlisting}[mathescape,language=easycrypt,xleftmargin=0pt,xrightmargin=0pt,style=easycrypt-pretty,basicstyle=\small]
theory SecretSharingScheme.
  const n : {int | 2 <= n} as n_pos.
  const t : {int | 0 <= t < n } as t_pos.
  type pid_t.
  op pid_set : pid_t list.

  type secret_t.
  type share_t.
  type shares_t = (pid_t * share_t) list.
  type rand_t.
  
  op share : rand_t -> secret_t -> (pid_t * share_t) list.
  op public_encoding : secret_t -> (pid_t * share_t) list.
  op pub_reconstruct : pid_t -> share_t -> secret_t.
  op reconstruct : (pid_t * share_t) list -> secret_t.
end SecretSharingScheme.
  \end{lstlisting}

The notion of $t$-privacy for a fixed share simulator $S$ is defined
by the following \textit{real} and \textit{ideal} evaluator.

\begin{lstlisting}[mathescape,language=easycrypt,xleftmargin=0pt,xrightmargin=0pt,style=easycrypt-pretty,basicstyle=\small]
module type Simulator_t = {
  proc simulate(aux : aux_t, cr : pid_t list) : shares_t
}.

module RealEvaluator = {
  proc share(aux : aux_t, r : rand_t, s : secret_t, cr : pid_t list) : shares_t = {
    var ss, ssc;
    ss <- share r s;
    ssc <- map (fun pid => (pid, oget (assoc ss pid))) cr;
    return ssc;
  }
}.

module IdealEvaluator (S : Simulator_t) = {
  proc share(aux : aux_t, r : rand_t, s : secret_t, cr : pid_t list) : shares_t = {
    var ssc;
    ssc <- S. simulate(aux,cr);
    return ssc;
  }
}.
\end{lstlisting}

A $t$-privacy claim for an $n$-party secret sharing scheme 
can be written in \EasyCrypt as follows.
\begin{lstlisting}[mathescape,language=easycrypt,xleftmargin=0pt,xrightmargin=0pt,style=easycrypt-pretty,basicstyle=\small]
  forall aux x,
  | Pr [ SSGame (D,R,RealEvaluator).main(aux,x) : res ] - 
        Pr [ SSGame (D,R,IdealEvaluator(S)).main(aux,x) : res ] | <= epsilon
\end{lstlisting}

\paragraph{Commitment schemes}
We close this section with a brief description
of our abstract formalization of commitment schemes,
which is essentially the same as that used in~\cite{priormpc}.

\begin{lstlisting}[mathescape,language=easycrypt,xleftmargin=0pt,xrightmargin=0pt,style=easycrypt-pretty,basicstyle=\small]
theory CommitmentScheme.  
  type msg_t.
  type rand_t.
  type commitment_t.
  type opening_string_t.
  type commit_info_t = commitment_t * opening_string_t.
  op commit : rand_t -> msg_t -> commit_info_t.
  op verify : msg_t -> commit_info_t -> bool.
end CommitmentScheme.
\end{lstlisting}

The \EasyCrypt definition of the correctness, binding and hiding properties
exactly match the definitions given in Section~\ref{sec:preliminaries} and
the follow the style of the definitions presented in this section.
We omit them for brevity.

\subsection{Formalizing the MPC-in-the-Head Transformation}

Our formalization of MPC-in-the-Head relies only on the
previous high-level abstractions and can be instantiated with
any MPC protocol, commitment scheme and secret sharing scheme
that meet the given syntax, correctness and security requirements.
We followed the modular structure of~\cite{MitH} in our formalization,
so the \ec{MPCInTheHead} theory below relies on sub-theories
for the underlying components.
Our result can be instantiated with any MPC protocol that
supports an arbitrary number of parties, but here we fix $n=5$,
as this allows us to explicitly unfold the hybrid arguments
that appear in the proof, which reduces proof complexity.


  \begin{lstlisting}[mathescape,language=easycrypt,xleftmargin=0pt,xrightmargin=0pt,style=easycrypt-pretty,basicstyle=\small]
theory MPCInTheHead.
  type witness_t. 
  type instance_t.
  clone import SecretSharingScheme as SS with 
    type secret_t <- witness_t, op n = 5, op t = 2.
  clone import MPCProtocol as MPC with  
    op n = SS.n, op t = SS.t,
    type pid_t = SS.pid_t, op pid_set = SS.pid_set,
    type pinput_t = instance_t,
    type sinput_t = share_t,
    type output_t = bool.
  clone import CommitmentScheme as CS with type msg_t = view_t.
  op relc : circuit_t.
  type statement_t = statement_instance_t.

  axiom good_circuit (x : statement_instance_t) w :
       valid_circuit relc /\ 
       forall ss ss', 
          exists r r', ss = share w r => ss' = share w r' =>
           f relc (mkseq (fun i => (x,(nth witness ss i).`2)) SS.n) = 
              f relc (mkseq (fun i => (x,(nth witness ss' i).`2)) SS.n).

  op relation(w : witness_t ,x : statement_instance_t) =
     exists (ss : (pid_t * share_t) list), 
        (w = SS.unshare ss /\
          let ins = mkseq (fun i => (x,(nth witness ss i).`2)) SS.n in
           let outs = f relc ins in
              all (fun o => o) outs).
  ...  
\end{lstlisting}

Note that the relation that is associated with the ZK protocol
is defined as the acceptance by all parties of the circuit 
computed by the MPC protocol. 
We bind the secret type of the secret sharing scheme to that of
witnesses and the input types of the MPC protocol to statements
and secret shares, as expected. 
Intuitively, it suffices that the prover is able to find a 
sharing of \ec{w} such that the MPC circuit accepts for the
relation to hold; naturally, we restrict our attention to 
circuits that are oblivious of which particular sharing is
chosen by the prover (axiom \ec{good\_circuit}).
However, the notion of circuit, witness and statement are still abstract.
The output type of the MPC protocol parties is set to \ec{bool} and
commitments are restricted to operate over protocol views.
Apart from these refinements, the other type definitions remain abstract
and can be instantiated arbitrarily.

The protocol itself instantiates the \ec{ZKProtocol} theory and fixes
the necessary types accordingly: the prover's commitment message is
a tuple of commitments corresponding to the views of the five
MPC protocol parties; the challenge returned by the verifier is
a pair of party identifiers, and the response returned by the
prover is a pair of opening strings for the selected views.

  \begin{lstlisting}[mathescape,language=easycrypt,xleftmargin=0pt,xrightmargin=0pt,style=easycrypt-pretty,basicstyle=\small]
  type commitment_t = (pid_t * CS.commitment_t) list. 
  
  type challenge_t = pid_t * pid_t.
  
  type response_t = (MPC.view_t * CS.opening_string_t) * (MPC.view_t * CS.opening_string_t).
\end{lstlisting}
  
We omit the details of the types of the randomness taken by
prover and verifier. In the case of the prover this includes
all the randomness required for secret sharing, emulating the 
MPC protocol and generating the commitments. 
The case of the verifier is simpler, since its randomness
includes only the choice of party identifiers for the challenge.

We give a short snippet of the commitment generation operator,
i.e., the first stage of the prover in the MitH construction,
where the prover emulates the MPC protocol execution and
commits to the view of party \ec{P1}. The full code replicates
the same process for all parties.

\begin{lstlisting}[mathescape,language=easycrypt,xleftmargin=0pt,xrightmargin=0pt,style=easycrypt-pretty,basicstyle=\small]
  op gen_commitment (rp : prover_rand_t) (xp : prover_input_t) : prover_state_t * commitment_t =
    let (w,x) = xp in
    let (c,x) = x in
    let (r_ss, r_mpc, r_cs) = rp in

    let ws = SS.share r_ss w in

    let x_mpc = map (fun pid => (pid, (x,oget (assoc ws pid)))) SS.pid_set in
    let (tr,y) = MPC.protocol c r_mpc x_mpc in

    let vs = map (fun pid => (pid, (input pid x_mpc, rand pid r_mpc, trace pid tr))) SS.pid_set in
    let cvs = map (fun pid => 
                    let r_c = oget (assoc r_cs pid) in
                    let v = oget (assoc vs pid) in 
                    (pid, (v, commit r_c v))) SS.pid_set in
    let cs = map (fun pid => (pid, fst (snd (oget (assoc cvs pid))))) SS.pid_set in
    (cvs, cs).
\end{lstlisting}

The challenge and response steps are simpler to formalize.
The challenge is simply a random sampling of a pair of identifiers,
which translates to copying random values from the randomness input.
The response selects the views and opening strings corresponding
to the selected parties, which are kept as internal state by the
prover using the \ec{get_party_committed_view}
operator. We give the \EasyCrypt code below.

  \begin{lstlisting}[mathescape,language=easycrypt,xleftmargin=0pt,xrightmargin=0pt,style=easycrypt-pretty,basicstyle=\small]
  op gen_challenge (rv : verifier_rand_t) (xv : verifier_input_t) (c : commitment_t) : verifier_state_t * challenge_t = ((rv,xv,c),rv).

  op gen_response (rp : prover_rand_t) (xp : prover_input_t) (c : commitment_t) (ch : challenge_t) (stp : prover_state_t) : response_t = 
    let cvs = stp in
    let (i,j) = ch in

    let cvi = get_party_committed_view i cvs in
    let (vi, cii) = cvi in
    let cvj = get_party_committed_view j cvs in
    let (vj, cij) = cvj in

    ((vi, snd cii), (vj, snd cij)).
\end{lstlisting}

Finally the verifier checks the response as follows.
  \begin{lstlisting}[mathescape,language=easycrypt,xleftmargin=0pt,xrightmargin=0pt,style=easycrypt-pretty,basicstyle=\small]
  op check (xv : verifier_input_t) (cs : commitment_t) (rv : verifier_rand_t) (r : response_t) : bool = 
    let (c,x) = xv in
    let (i,j) = rv in

    let (vosi, vosj) = r in
    let (vi, osi) = vosi in
    let (vj, osj) = vosj in

    let (xi,ri,tri) = vi in
    let (xj,rj,trj) = vj in

    let ci = get_party_commitment i cs in
    let cj = get_party_commitment j cs in

    CS.verify vi (ci,osi) /\ CS.verify vj (cj,osj) /\
    MPC.consistent_views c x vi vj i j /\
    MPC.local_output c i (xi,ri,tri) /\ MPC.local_output c j (xj,rj,trj).
\end{lstlisting}

The \ec{check} operator verifies three conditions: 
1) that the
commitment openings are valid wrt to the provided views;
2) the provided views are consistent with
each other (using operator \ec{consistent_views}); and
3) that the local output reported by the
selected parties is \textit{true}, which implies that the
MPC protocol execution for these parties reported that
the relation between statement and witness holds.

\paragraph{Completeness}
Our completeness theorem  states that
the MitH construction has perfect completeness assuming perfect correctness
for the underlying components. Formally, in \EasyCrypt we prove that for 
all valid randomness samplers \ec{R}, all statements \ec{x} and all
witnesses \ec{w}, the following holds.

  \begin{lstlisting}[mathescape,language=easycrypt,xleftmargin=0pt,xrightmargin=0pt,style=easycrypt-pretty,basicstyle=\small]
lemma completeness w x:
   relation w x => Pr [ Compleneness(R).main(w,x) : res ] = 1%r.
\end{lstlisting}

  The proof intuition is as follows.
  The \ec{good\_circuit} restriction imposes that the circuit 
  that defines the relation is well
  behaved, in the sense that, for all sharings 
  $\bar{w},\bar{w}' \in [\Share(w)]$, the circuit
  outputs the same consistent values for all parties.
  Then, if the MPC protocol is
  correct, it will correctly compute the relation of the ZK
  proof system and every two
  views will be pairwise consistent (by Lemma~\ref{th:consistency}),
  Since this is an honest execution, the commitments are well constructed
  and the openings will be valid, as per the correctness property of
  the commitment scheme.

\paragraph{Soundness}

Our MitH implementation is sound for a soundness error of $1 - 1/\binom{n}{2} +
\epsilon$ in a single execution. Here $n$ is the number of parties in
the MPC protocol and $\epsilon$ is bounded using the binding property
of the commitment scheme. 
The statement in \EasyCrypt is as follows:

\begin{lstlisting}[mathescape,language=easycrypt,xleftmargin=0pt,xrightmargin=0pt,style=easycrypt-pretty,basicstyle=\small]
lemma soundness xc (MP <: MProver_t) :  
   !language (snd xc) =>
   Pr [ Soundness(RV,MP).main(x) : res ] <= 2%r * Pr [Binding(A).main() : res ] + (1 - $1/\binom{n}{2}$)
\end{lstlisting}

  The proof is done by a sequence of two game hops.
  In the first hop, we specify a bad event that checks if the
  dishonest prover opened a commitment for the first  view
  requested in the challenge that is not
  the originally committed one and reduce this bad event
  to the binding property of the commitment scheme.
  The second hop repeats the reduction, but for the second
  opened view. 
  Finally we, prove that the probability that the prover
  cheats in the resulting game is $0$ bounded by 
  $1/\binom{n}{2}$. 
  Intuitively, if the verifier accepts then both
  views are consistent with each other and they
  both return $1$. 
  However, we know that no sharing of $w$ would cause
  the circuit to make all parties return $1$ in an honest
  execution.
  Here we rely on Lemma~\ref{th:consistency}, which we axiomatize in \EasyCrypt
  as follows.
  
  \begin{lstlisting}[mathescape,language=easycrypt,xleftmargin=0pt,xrightmargin=0pt,style=easycrypt-pretty,basicstyle=\small]
axiom local_global_consistency (c : circuit_t) (xp : pinput_t list) (vv : view_t list) :
  (forall i j, consistent_views c xp (get_view i vv) (get_view j vv) i j) <=>
  (exists xs rs, let (tr,y) = eval_protocol c rs (xp,xs) in
                       forall i, ((pinput i xp, sintput j xs), get_rand i rs, get_trace i tr) = get_view i vv   get_view i.
\end{lstlisting}

  This states that there must exist a pair of views that fails the checks
  performed by the verifier. 
  The soundness result follows from the fact that the (honest)
  verifier chooses two views at random for opening, so the probability
  of hitting one problematic pair of views is at least
  $1/\binom{n}{2}$

\paragraph{Zero-knowledge}
For the Zero-Knowledge property we formalize a simulation strategy
that guesses the challenge the verifier will produce
by generating it uniformly at random.
It then runs the MPC simulator to generate the two views that
will be opened and fixes the other ones to an arbitrary value.
Note here that the outputs of the simulated views can be programmed
to $1$.
It completes these views by sampling two random shares using
the secret sharing scheme's share simulator.
Finally, it commits to all views to get a simulated first round
message.
When computing the response, this simulation strategy fails
if the challenge guess was wrong. Otherwise it returns the simulated
views.

  The proof that this is a good simulation strategy is established as
  two independent results. The first result shows that a single execution
  of our simulation strategy is good if the simulator's guess is
  successful. The second result is a meta-theorem described at the
  end of this section.

  To prove the fist result, we first define a modified real game 
  that initially samples a 
  challenge uniformly at random rand aborts if the verifier's
  challenge does not equal the randomly sampled one.
  We then prove, using a sequence of hops, that a single run
  of our simulation strategy is indistinguishable from this
  modified real game for any distinguisher
  that only observes it if the challenge guess is sucessful.
  This is stated as follows, where \ec{RealEvaluatorMod}
  denotes the modified real world.

  \begin{lstlisting}[mathescape,language=easycrypt,xleftmargin=0pt,xrightmargin=0pt,style=easycrypt-pretty,basicstyle=\small]
lemma zero_knowledge w x (D <: Distinguisher_t) (MV <: MVerifier_t) :  
   valid_circuit (fst xc) =>
   | Pr [ ZKGame(D,RP,RealEvaluatorMod(MV)).game(w,x) : res ] -
     Pr [ ZKGame(D,RP,IdealEvaluator(MV,S(S$_{\mathsf{MPC}}$))).game(w,x) : res ] |<=
     3%r $\cdot$ | Pr [ Hiding(B(D)).game(true) : res ] - Pr [ Hiding(B(D)).game(false) : !res ] | +
     | Pr [ PrivGame[(D'(D),R$_{\mathsf{MPC}}$,RealEvaluator).main(fst x,w,cr) : res ] -
       Pr [ PrivGame[(D'(D),R$_{\mathsf{MPC}}$,IdealEvaluator(S$_{\mathsf{MPC}}$)).main(fst x,w,cr) : res ] | +
     | Pr [ SSGame (C(D),R$_{\mathsf{SS}}$,RealEvaluator).main(aux,x) : res ] - 
        Pr [ SSGame (C(D),R$_{\mathsf{SS}}$,IdealEvaluator(S)).main(aux,x) : res ] |
\end{lstlisting}

  We now detail the game hopping proof.
  In the first three hops, we replace the view of every party
  different from $i$ and $j$, where $(i,j)$ randomly sampled
  challenge, by the default \EasyCrypt value
  \ec{witness}. We can construct adversary \ec{B} that attacks
  the hiding property of the commitment scheme and interpolates
  between the two experiments as follows. Adversary \ec{B} executes
  the commitment round in the exact same way as the protocol does in
  the real world. It then selects one party (different from $i$ and
  $j$) and adopts as its query the view of that party and the
  \ec{witness} value. When \ec{B} finishes the execution
  of the MitH protocol, it adopts as its decision bit the one given by
  the zero-knowledge distinguisher. These three hops lead to a 
  term in the final statement that is bounded by $3$ times the
  advantage of \ec{B} against the commitment scheme.
  Next, we replace the execution of the MPC protocol by its simulator. 
  Because, by assumption, the MPC protocol achieves
  $2$-privacy, it is possible to bound the difference between the two
  games as the advantage of a distinguisher against the privacy of the
  MPC protocol.
  Finally, in the last hop, we replace the value of the witness $w$ by
  an arbitrary value and reduce the transition to the $t$-privacy
  of the secret-sharing scheme. The proof is complete by showing
  that this final game is identical to the ideal game when instantiated
  with our simulator.

\paragraph{Meta Theorems} We have proved in \EasyCrypt three meta-theorems
that allow us to justify why our soundness and ZK results imply the
standard soundness properties (wich negligible soundness error) and
ZK property. 
The first meta-result shows that $n$ repetitions of an experiment that
returns $1$ with small probability $\epsilon$, and checking that all
attempts succeeded has an overall probability of success bounded by 
$\epsilon^n$.
The second meta-result considers an algorithm such as our simulator
that is good enough when its challenge guess is correct.
It then shows that rejection sampling of such a simulator permits
obtaining a computationally indistinguishable simulation that
only fails with negligible probability.
Finally, the third meta theorem shows that any protocol that
follows the next message syntax satisfies the global consistency
versus pairwise consistency property.


\section{Verified ZKP implementation supporting general relations}
\label{sec:verif-eval-engine}

We now describe how the abstract MitH construction
was instantiated to derive a concrete
formalization of a MitH protocol based on the Shamir's secret sharing
scheme, the BGW protocol~\cite{ben1988completeness} and the Pedersen 
commitment scheme.
We leveraged prior results in the verification of MPC protocols~\cite{priormpc}, 
including a formalization of these lower 
level components and a code extraction tool that 
automatically synthesizes OCaml code from \EasyCrypt specifications.
As a result, we obtain a verified implementation of our
MitH instantiation, with only small adaptations to the original formalization and extraction tool.
To demonstrate the modularity of our development, we also replace the
Pedersen commitment scheme by a more efficient PRF-based construction.
We conclude the section with a preliminary performance
analysis of our MitH verified implementation.

\subsection{Secure arithmetic circuit evaluation}


As a first step, we refine the notion of MPC protocol to the concrete
case of secure arithmetic circuit evaluation where parties evaluate
addition, multiplication and scalar multiplication gates sequentially.
The \EasyCrypt definition of an arithmetic circuit is depicted below. 

\begin{lstlisting}[mathescape,language=easycrypt,xleftmargin=0pt,xrightmargin=0pt,style=easycrypt-pretty,basicstyle=\small]
theory ArithmeticCircuit.
  type wire_t = t.
  type gid_t = int.
  type topology_t = int * int * int.
  
  type gates_t = [
    | PInput of int
    | SInput of int
    | Constant of int & t
    | Addition of int & gates_t & gates_t
    | Multiplication of int & gates_t & gates_t
    | SMultiplication of int & gates_t & gates_t
  ].

  type circuit_t = topology_t * gates_t.  
end ArithmeticCircuit.
  \end{lstlisting}

The circuit is defined over wires, which are elements 
of a finite field \ec{t}. 
The topology (type \ec{topology_t}) is a pair of integers
that fixes the number of public input wires, the number
of secret input wires and the number of gates to the circuit.
Intuitively, when $n$ parties securely evaluate such a 
circuit, all of them will receive the values of the
public input wires in the clear, which is consistent with
our assumption in Section~\ref{sec:mpccomponent} that
all parties receive the same public input.
The \ec{gates_t} permits specifying an arbitrary 
arithmetic circuit using a tree structure. 
This is less efficient that a graph structure, but it
reduces the verification effort.

From this point on, different protocols
for secure arithmetic circuit evaluation can be obtained by
instantiating the three gate-level secure computation sub-protocols.
Indeed, our refinement of the MPC protocol formalization is
parametric on the specification of these gates.%
\footnote{We show an example
of such an arithmetic gate specification in Appendix~\ref{app:abstr-arithm-gate}.} 
However, we make concrete the number of parties $n=5$, 
as this is all we need for the concrete setting where
we use the BGW protocol instantiation to instantiate
the MitH construction.
Note that we define the trace of messages to follow the same 
structure as that of the circuit, which again simplifies proofs.

\begin{lstlisting}[mathescape,language=easycrypt,xleftmargin=0pt,xrightmargin=0pt,style=easycrypt-pretty,basicstyle=\small]
theory ArithmeticProtocol.
  clone import AddGate.
  clone import MulGate.
  clone import SMulGate.
  op n = 5. op t = 2.
  op pid_t = [ | P1 | P2 | P3 | P4 | P5 ].
  type pinput_t = t list. 
  type sinput_t = t list.
  type output_t = t.
  type randg_t = [
    | AddRand of AddGate.rand_t
    | MulRand of MulGate.rand_t
    | SMulRand of SMulGate.rand_t
  ].
  type rand_t = (gid_t * randg_t) list.
  type msgs_t = [
    | PInputM of wid_t
    | SInputM of wid_t
    | ConstantM of gid_t & t
    | AdditionM of gid_t & AdditionGate.Gate.msgs_t & msgs_t & msgs_t
    | MultiplicationM of gid_t & MultiplicationGate.msgs_t & msgs_t & msgs_t
    | SMultiplicationM of gid_t & SMultiplicationGate.msgs_t & msgs_t & msgs_t
...
\end{lstlisting}

At this level of abstraction, we can define the secure evaluation of a
circuit using a secret-sharing based protocol which captures the BGW
protocol as a particular case. The remaining details are fixed by
instantiating the secret sharing scheme and the arithmetic gates.
Note below that the \ec{input_t} type for secret inputs 
is defined as a share of an initial input value and the. Moreover, the
output of the protocol is computed by explicitly 
taking the partial outputs produced by each party and using
them to reconstruct an unshared output value that is then
fixed as the output for all parties as a  consequence of
a public opening. This choice of design comes from a security
relaxation that is made at the circuit
evaluation level, prior to the output phase. To obtain a $t$-privacy
arithmetic protocol, it is still necessary to compose the
protocol provided here with a gate that securely implements a share randomization
functionality. This compositional property is studied in
\cite{bogdanov2008sharemind,almeida2018enforcing} and 
its formalization was adapted from \cite{priormpc}. 

  \begin{lstlisting}[mathescape,language=easycrypt,xleftmargin=0pt,xrightmargin=0pt,style=easycrypt-pretty,basicstyle=\small]
  ...
  op eval_circuit (gg : gates_t) (r : rand_t list) (x : input_t list) : trace_t * output_t =
    ...
    with gg = SInput w => 
      let ys = map (fun pid => let sec = nth witness (sinput pid xs) w in (pid, sec)) pid_set in
      let tr = SInputT w in (tr, ys)
    with gg = Addition gid wl wr => 
      let ra = map (fun pid => (pid, as_gate_rand_addition (oget (assoc (oget (assoc rs pid)) gid)))) pid_set in
      let (tl, vwl) = eval_gates wl rs xs in
      let (tr, vwr) = eval_gates wr rs xs in
      let gxs = map (fun pid => (pid, ((), (oget (assoc vwl pid), oget (assoc vwr pid))))) pid_set in
      let (gtr, gys) = AdditionGate.Gate.gate ra gxs in
      let gtrs = AdditionT gid gtr tl tr in  (gtrs, gys)
    with gg = Multiplication gid wl wr => ...
    with gg = SMultiplication gid wl wr => ...

  op protocol (c : circuit_t) (r : rand_t list) (x : input_t list) : trace_t * outputs_t list =
    ...
    let cc = snd c in
    let c_out = eval_circuit cc r x in
    let tr' = unzip1 c_out in 
    let y' = unzip2 c_out in
    let y = map (fun y => oget (reconstruct y)) y' in 
    let tr = (y',tr') in (tr, (y,y,y,y,y)).
  ...
\end{lstlisting}

We end the discussion of our arithmetic protocol specification with
the refinement of the \ec{local_output} and \ec{out_messages}
operators. Similarly to the protocol evaluation function, these
two operators are defined based on the operators with the same
names defined by the lower-level arithmetic gate theories.

  \begin{lstlisting}[mathescape,language=easycrypt,xleftmargin=0pt,xrightmargin=0pt,style=easycrypt-pretty,basicstyle=\small]
  ...
  op local_output_gates (pid : pid_t) (tr : trace_t) (x : pinput_t * sinput_t) (r : rand_t) : output_t =
    with im = PInputLT w => ...
    with im = SInputLT w => nth witness (snd x) w
    with im = ConstantLT gid c => ConstantGate.local_output pid ((c,()), (), [])
    with im = AdditionLT gid tr trl trr => 
      let vl = local_output_gates pid x r trl in
      let vr = local_output_gates pid x r trr in
      let ra = as_gate_rand_addition (oget (assoc r gid)) in
      AdditionGate.local_output pid (((),(vl,vr)), ra, tr)
    with tr = MultiplicationLT gid ta tl tr => ...
    with tr = SMultiplicationLT gid ta tl tr => ...
  op local_output (c : circuit_t) (pid : pid_t) (v : view_t) : output_t =
    let (topo, gg) = c in let (x,r,tr) = v in
    if valid_circuit_trace c tr then local_output_gates pid x r tr  else witness.

  op gate_out_messages (pid : pid_t) (t : gate_local_trace_t) (x : input_t) (r : rand_t) : gate_local_trace_t =
    with im = PInputLT w => PInputLT w
    with im = SInputLT w => SInputLT w
    with im = ConstantLT gid c => ConstantLT gid c
    with im = AdditionLT gid tr trl trr => 
      let vl = local_output_gates pid x r trl in
      let vr = local_output_gates pid x r trr in
      let owl = out_messages_gates pid x r trl in
      let owr = out_messages_gates pid x r trr in
      let ra = as_gate_rand_addition (oget (assoc r gid)) in
      AdditionLT gid (AdditionGate.out_messages pid (((),(vl,vr)), ra, tr)) owl owr
    with t = MultiplicationLT gid ta tl tr => ...
    with t = SMultiplicationLT gid ta tl tr => ...
  op out_messages (c : circuit_t) (pid : pid_t) (v : view_t) : out_messages_t =
    let (topo, gg) = c in let (x,r,tr) = v in
    if valid_circuit_trace c tr then out_messages_gates pid x r tr else witness.
end ArithmeticProtocol.
  \end{lstlisting}

The BGW protocol is obtained by instantiating the secret sharing
scheme with Shamir's secret sharing and the low-level arithmetic
gate protocols (including the randomization output gate
\textit{refresh}) proved secure in~\cite{priormpc}.
We briefly describe how correctness and security of the full
protocol are proved in \EasyCrypt.
Correctness is proved by induction on the structure of the
circuit and relying on the correctness of the low-level gates
at each inductive step.
The $2$-privacy of the protocol is provide by instantiating the
secure composition theorem, assuming that the low-level
arithmetic gates are $t$-pre-output-privacy and that the refresh gate is $t$-privacy.

\begin{lstlisting}[mathescape,language=easycrypt,xleftmargin=0pt,xrightmargin=0pt,style=easycrypt-pretty,basicstyle=\small]
lemma privacy_arithmetic (D <: Distinguisher_t) c x aux ,
  valid_secret x =>
  | Pr [ PrivGame(D,R,RealEvaluator).main(c,x,aux) : res ] -
    Pr [ PrivGame(D,R,IdealEvaluator(S)).main(c,x,aux) : res ] | <=
     | Pr [ PreOutPrivGame$^{\mathsf{ArithProt}}$(D$_{\mathsf{ArithProt}}$(D),R$_{\mathsf{ArithProt}}$,RealEvaluator).main($\mathsf{ArithProt}$,x,aux) : res ] - 
      Pr [ PreOutPrivGame$^{\mathsf{ArithProt}}$(D$_{\mathsf{ArithProt}}$(D),R$_{\mathsf{ArithProt}}$,IdealEvaluator(S$_{\mathsf{ArithProt}}$)).main($\mathsf{ArithProt}$,x,aux) : res ] | +
        | Pr [ PrivGame$^{\mathsf{Refresh}}$(D$_{\mathsf{Refresh}}$(D),R$_{\mathsf{Refresh}}$,RealEvaluator).main($\mathsf{Refresh}$,x,aux) : res ] - 
         Pr [ PrivGame$^{\mathsf{Refresh}}$(D$_{\mathsf{Refresh}}$(D),R$_{\mathsf{Refresh}}$,IdealEvaluator(S$_{\mathsf{Refresh}}$)).main($\mathsf{Refresh}$,x,aux) : res ] |
\end{lstlisting}

\noindent where \ec{PreOutPrivacy} represents the $t$-privacy
relaxed security definition used at the protocol evaluation level.

\begin{proof}

  The proof is done in two hops, where the executions of the low-level 
  $t$-pre-output-privacy secure arithmetic gate evaluation
  protocol and the $t$-privacy refresh gate are replaced by their
  corresponding simulators. For example, it is possible to define \textit{real} evaluator
  \ec{RealEvaluator1} that replaces every execution of
  the arithmetic protocol by simulator \ec{S}$_{\mathsf{ArithProt}}$ and
  bounding the difference to the real world  by building an
  adversary against $\mathsf{ArithProt}$. When the execution of the
  composed protocol is replaced by the respective simulator, we will be in the
  ideal world and it is possible to
  trace back the difference between \ec{RealEvaluator} and
  \ec{IdealEvaluator} to be the sum of the 
  advantages against the arithmetic protocol and of the refresh gate.
  
  
\end{proof}

\subsection{PRF-based commitment scheme}

Our first realization of the MitH construction
considered the Pedersen commitment scheme in the computation of
commitments for the party views resulting from the protocol
execution. This particular commitment scheme works just for
one finite field element, which means that  
producing a commitment to a view---a large list of finite field 
elements---implies producing a
commitment to every value in the view.
For this reason, we formalized the (collision-resistant) 
PRF-based construction given
in~\cite{hmaccommitment}, which we then instantiate with HMAC.
To this end, we first formalize the underlying PRF primitive as a keyed function as follows.

  \begin{lstlisting}[mathescape,language=easycrypt,xleftmargin=0pt,xrightmargin=0pt,style=easycrypt-pretty,basicstyle=\small]
theory CRPRF.
  type input.
  type output.

  type key.

  op f : key -> input -> output.
end CRPRF.    
\end{lstlisting}

Using a PRF that follows this syntax, it is possible to build a
commitment scheme by defining the commit algorithm to be an evaluation
of the PRF and the verifying algorithm to be the comparison between
the received commitment and the one that the verifying party is able
to locally compute once the PRF key is revealed.

  \begin{lstlisting}[mathescape,language=easycrypt,xleftmargin=0pt,xrightmargin=0pt,style=easycrypt-pretty,basicstyle=\small]
theory CRPRFCommitment.
  clone include CRPRF.
  op commit(k,m) = (f k m,k).
  op verify(m : input, ci : output * key) = ci.`1 = f ci.`2 m.
  clone import CommitmentScheme with
    type msg_t = input,
    type rand_t = key,
    op valid_rand =  fun _ => true,
    type opening_string_t = key,
    type commitment_t = output,
    op commit = commit,
    op verify = verify.
  \end{lstlisting}

Our binding theorem below states that our PRF-based commitment scheme
achieves binding if the underlying PRF is collision resistant.

\begin{lstlisting}[mathescape,language=easycrypt,xleftmargin=0pt,xrightmargin=0pt,style=easycrypt-pretty,basicstyle=\small]
lemma prf_commitment_binding (A <: Adversary_t) :
  Pr [ Binding(A).main() : res ] <= Pr [ CR$_{\mathsf{PRF}}$(B(A)).main() : res ] 
\end{lstlisting}

\noindent where \ec{Pr[ CR().main() : res ]}
represents the probability of finding a collision under \ec{PRF}.

\begin{proof}

  The proof is a classical reduction to the collision resistance
  property. Based on adversary \ec{A}, we are able to construct
  adversary \ec{B} that works as 
  follows. To produce its query, \ec{B} invokes the query generation
  adversary from \ec{A} and adopts its query as its own. The 
  collision resistance game will then execute in the exact same way as
  the binding game, since in both experiences the challenger will
  check if the two queried values are different and if they produce
  the same commitment.
  
\end{proof}

Finally, we prove that our PRF-based commitment scheme achieves
hiding.

\begin{lstlisting}[mathescape,language=easycrypt,xleftmargin=0pt,xrightmargin=0pt,style=easycrypt-pretty,basicstyle=\small]
lemma prf_commitment_hiding (A <: Adversary_t) : 
  | Pr [ Hiding(A).game(true) : res ] - Pr [ Hiding(A).game(false) : !res ] | <=
    1/2 + | Pr [ PRF(B(A)).game(true) : res ] - Pr [ PRF(B(A)).game(false) : !res ] |
\end{lstlisting}

\noindent  where \ec{| Pr[ PRF(B(A)).game(true) : res] - Pr[
  PRF(B(A)).game(false) : !res] |} represents the advantage against
the underlying PRF assumption.

\begin{proof}

  The proof is a classical reduction to the PRF assumption. Based on
  adversary \ec{A}, we are able to construct adversary \ec{B} against
  the PRF assumption that works as follows. To produce its query,
  \ec{B} invokes the query generation method from adversary \ec{A} and
  adopts the obtained query as its own. To make its guess, \ec{B} will
  simply adopt the decision bit from \ec{A} as its own. Note that, in
  the process, \ec{B} has access to an oracle that computes the
  outputs of the PRF.

  Finally, because of the PRF assumption, we can replace the output of
  \ec{f} by a random output value, giving the adversary a
  \ec{1 / 2} probability on winning the game.
  
\end{proof}

\subsection{MitH-based verified evaluation engine for ZKP}

We have obtained a verified implementation of our MitH \EasyCrypt
implementation via the \EasyCrypt
extraction tool developed in \cite{priormpc}, that was refined to be used in the
particular context of this work. This
general-purpose tool is able to synthesize correct-by-construction
executable code from an \EasyCrypt proof script. 

We have applied the extraction tool at two different
levels. First, we applied the extraction tool against a concrete
\EasyCrypt specification of MitH, where commitments were computed using
the Pedersen commitment scheme. 
Nevertheless, this extraction strategy loses the modular
infrastructure that surrounds the proof and relies on a commitment
scheme that entails a significant performance penalty (a performance
analysis can be consulted in Table~\ref{tab:mith_perf}). Therefore,
after specifying the efficient commitment scheme based on a PRF,
we applied a different code extraction approach to it:
instead of performing a \textit{flat} code synthesis that disregarded
the modular structure, we manually extracted the modular structure and
automatically extracted the concrete components, that were later matched
against the modular structure. This methodology allowed us to a
modular implementation of MitH which is re-usable and more efficient
because it does not rely on group operations to compute the
required commitments.

\begin{figure}[ht!]
  \centering
  \includegraphics[width=.7\textwidth]{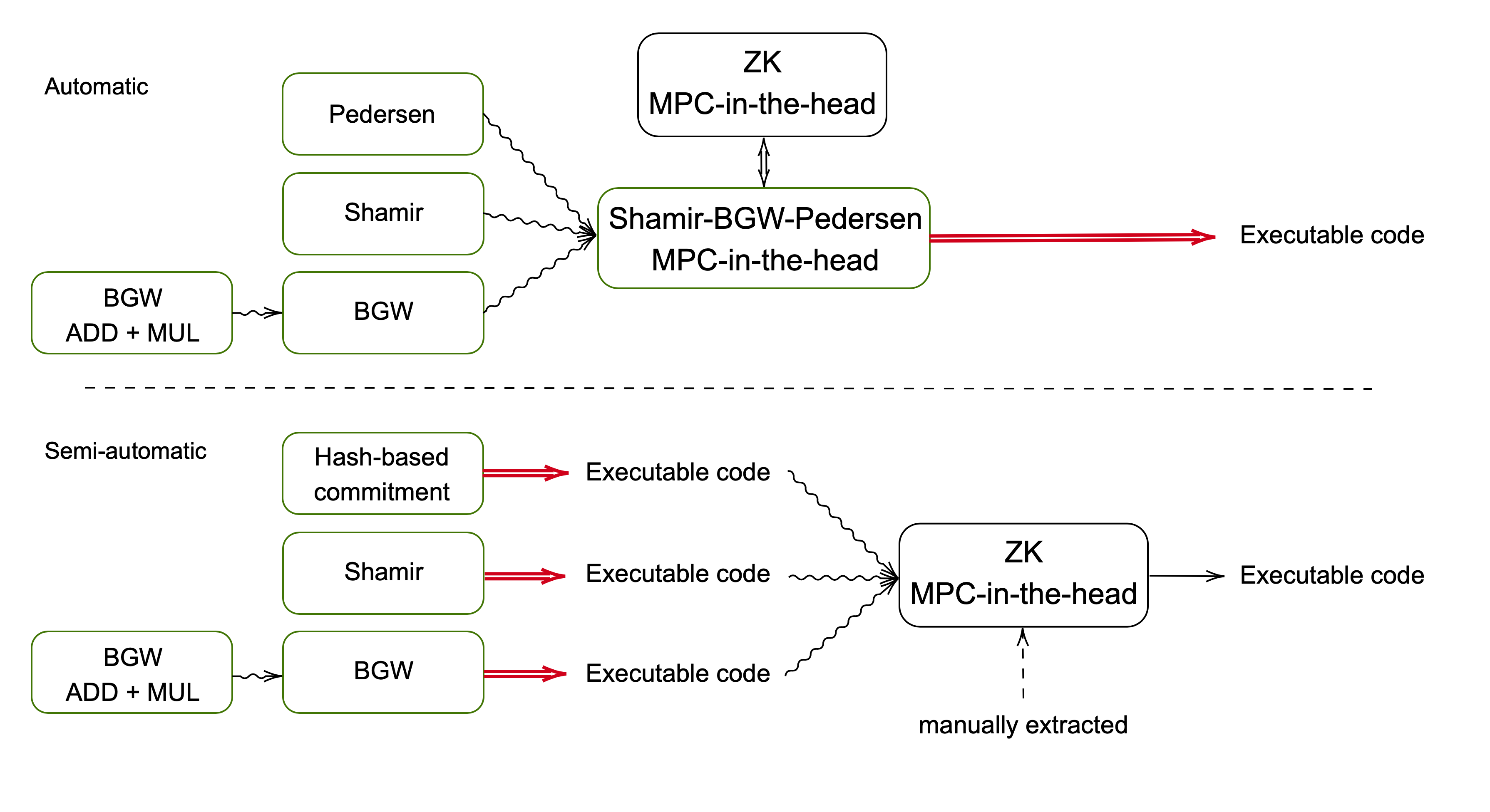}
  \caption{Code extraction pathways. The red arrow
    (\textcolor{red}{$\Rightarrow$}) denotes executable code 
    obtained via the automated extraction mechanism.}
  \label{fig:extraction-pathways}
\end{figure}

A graphical description of these two extraction pathways can be found
in Figure~\ref{fig:extraction-pathways}, where the former
extraction path is depicted on top and the latter is depicted
on the bottom.

\subsection{Experimental results}

We performed a preliminary performance analysis of our executable
\OCaml MitH implementation 
obtained from \EasyCrypt, that we summarize in
Table~\ref{tab:mith_perf}. The table contains not only the current
performance analysis of the MitH implementation, but also of the
individual cryptographic primitives that we developed, with different
field sizes. The benchmarking was carried out in a
commodity 13-inch 2016 MacBook Pro, with a dual-core Intel Core i5
processor clocked at 2.9 GHz, 16 GB RAM, 256 KB L2 cache and 4 MB L3
cache.

\begin{table}[ht!]
  \centering
    \caption{Performance analysis of automatically extracted
      executable code  for MitH-based ZKP (times in \textit{ms}).}
  \resizebox{\linewidth}{!}{%
\begin{tabular}{cc|cccccc}
\multicolumn{1}{c|}{}                                 &                           & Random generation   & Share                & Reconstruct          & Protocol             & Commit               & Verify               \\ \hline
\multicolumn{1}{c|}{\multirow{5}{*}{Field: 256 bits}} & Shamir Secret Share       & 0.008                & 0.010               & 0.002                & -                    & -                    & -                    \\
\multicolumn{1}{c|}{}                                 & BGW Addition              & -                    & -                    & -                    & 0.001                & -                    & -                    \\
\multicolumn{1}{c|}{}                                 & BGW Scalar Multiplication & -                    & -                    & -                    & 0.002                & -                    & -                    \\
\multicolumn{1}{c|}{}                                 & BGW Multiplication        & 0.028                & -                    & -                    & 0.053              & -                    & -                    \\
\multicolumn{1}{c|}{}                                 & Pedersen
                                                        Commitment
                                                                                  & 0.002                & -                    & -                    & -                    & 0.033               & 0.033               \\
  \multicolumn{1}{c|}{}                                 & SHA256-based Commitment       & -                & -                    & -                    & -                    & 0.003               & 0.002               \\ \hline
                                                      &                           &                      &                      &                      &                      &                      &                      \\ \hline
\multicolumn{1}{c|}{\multirow{5}{*}{Field: 1024 bits}}     & Shamir Secret Share       & 0.006                & 0.009                & 0.001                & -                    & -                    & -                    \\
\multicolumn{1}{c|}{}                                 & BGW Addition              & -                    & -                    & -                    & 0.002                & -                    & -                    \\
\multicolumn{1}{c|}{}                                 & BGW Scalar Multiplication & -                    & -                    & -                    & 0.004                & -                    & -                    \\
\multicolumn{1}{c|}{}                                 & BGW Multiplication        & 0.022                & -                    & -                    & 0.052                & -                    & -                    \\
\multicolumn{1}{c|}{}                                 & Pedersen
                                                        Commitment
                                                                                  & 0.002                & -                    & -                    & -                    & 1.093               & 1.090              \\
      \multicolumn{1}{c|}{}                                 & SHA256-based Commitment       & -                & -                    & -                    & -                    & 0.004               & 0.004               \\ \hline
\multicolumn{1}{l}{}                                  & \multicolumn{1}{l|}{}     & \multicolumn{1}{l}{} & \multicolumn{1}{l}{} & \multicolumn{1}{l}{} & \multicolumn{1}{l}{} & \multicolumn{1}{l}{} & \multicolumn{1}{l}{} \\ \hline
\multicolumn{1}{c|}{Field value: 101}                       &
                                                               \textbf{\textcolor{red}{A}}
                                                              MitH (7 gates, 2 MUL)   & 0.233                & -                    & -                    & 2.692                & \textbf{1.687}                   & \textbf{0.434}                   \\
\multicolumn{1}{c|}{Field value: 97}                        &
                                                              \textbf{\textcolor{red}{A}}  MitH (11 gates, 3 MUL)  & 0.396                & -                    & -                    & 3.946                & \textbf{1.962}                    & \textbf{0.520}                    \\ \hline
\multicolumn{1}{c|}{}                                 &                           &                      &                      &                      &                      &                      &                      \\ \hline
\multicolumn{1}{c|}{Field value: 101}                       & \textbf{\textcolor{blue}{SA}} MitH (7 gates, 2 MUL)   & 0.137                & -                    & -                    & 0.925                & \textbf{0.051}                  & \textbf{0.014}                  \\
\multicolumn{1}{c|}{Field value: 97}                        & \textbf{\textcolor{blue}{SA}} MitH (11 gates, 3 MUL)  & 0.175                & -                    & -                    & 1.166               & \textbf{0.077}                   & \textbf{0.029}                 
\end{tabular}}

  \label{tab:mith_perf}
\end{table}

As expected, our performance results are intrinsically related to the
field size that is used for the field operations. Moreover, we refer
the reader to the last two rows of the performance table, where is
possible to compare the MitH-based ZK implementation that
uses the Pedersen commitment scheme (\textbf{\textcolor{red}{A}}) and the implementation that uses
a PRF-based commitment scheme (\textbf{\textcolor{blue}{SA}}), instantiated with the SHA256 hash
function. Naturally, there are significant efficiency gains when using
the PRF-based commitment to the detriment of Pedersen commitment
scheme, mainly because the latter heavily relies on  group
operations, whereas the former is simply an application of an
(efficient) hash function. Finally, a rough comparison with results
for other implementations of the same primitives and of MitH-based ZK allows us
to conclude that the performance penalty induced by our formal
approach is not prohibitive and that real-world applications are
within reach of the implementations automatically generated by our
approach. Furthermore, additional optimization effort can lead to
significant performance gains, e.g., by resorting faster MPC
protocols that do not require polynomial interpolation (such as the
one proposed in \cite{maurer2006secure}) or by deploying highly
optimized low-level implementations of the cryptographic primitives
that we developed in this project, such as the ones given by the
Jasmin language.



\bibliography{refs}


\appendix

\section{Abstract arithmetic gate example}
\label{app:abstr-arithm-gate}
To show how arithmetic gates were formalized, we will provide, as
example, a walkthrough of how addition gates are abstracted as
sub-protocols in our formalization.

The addition gate works over elements of a finite field. Every party
enters the gate evaluation protocol with two input values, i.e., every
party holds a share of two values. The functionality will be given the
reconstruction of those two secrets from where the input shares come
from. All parties end the gate evaluation with one share of the output
value.

Because we are specifying a gate for a concrete operation, we can
already refine the functionality to be the addition operation over the
finite field applied to the field elements reconstructed from the
local inputs of the parties.

\begin{lstlisting}[mathescape,language=easycrypt,xleftmargin=0pt,xrightmargin=0pt,style=easycrypt-pretty,basicstyle=\small]
theory AdditionGate.
  op n : int.
  type pid_t = int.

  type pinput_t. 
  type sinput_t = t * t.
  type input_t = pinput_t * sinput_t.
  type output_t = t.
  op f  (xs : input_t list) : output_t list = map (+) xs.

  type rand_t.
  type nextmsg_t.
  type view_t = input_t * rand_t * nextmsg_t list.

  op out_messages : pid_t -> view_t -> (nextmsg_t list) option.
  op local_output : pid_t -> view_t -> output_t option.

  type trace_t = view_t list.
  op gate : input_t list -> rand_t list -> trace_t * output_t list.

end AdditionGate.
\end{lstlisting}

We purposely leave the randomness type
underspecified to capture different instantiations of addition
gates that may or may not require some random tape to securely perform
the required computations.

Note that all methods related with the actual gate evaluation are left
underspecified, meaning that we do not provide an actual realization
of an addition gate. Indeed, the purpose of this gate specification
was not to provide any concrete realization of an addition gate
but to lay the foundations for the future definition of secure
addition gate instantiations.

\end{document}